\def\year{2021}\relax
\documentclass[letterpaper]{article} 
\usepackage{aaai21}  
\usepackage{times}  
\usepackage{helvet} 
\usepackage{courier}  
\usepackage[hyphens]{url}  
\usepackage{graphicx} 
\usepackage{natbib}  
\usepackage{caption} 
\usepackage{enumitem}
\usepackage{amsmath}
\usepackage{graphicx}
\graphicspath{ {./images/} }
\title{JEL: Applying End-to-End Neural Entity Linking in JPMorgan Chase}
\author {
        Wanying Ding,
        Vinay K. Chaudhri,
        Naren Chittar,
        Krishna Konakanchi \\
}
\affiliations {
    JPMorgan Chase  \& Co. \\
    \{wanying.ding, vinay.chaudhri, naren.chittar\}@jpmchase.com, krishna.k.konakanchi@chase.com
}

\begin{document}

\maketitle

\begin{abstract}
Knowledge Graphs have emerged as a compelling abstraction for capturing key relationship among the entities of interest to enterprises and for integrating data from heterogeneous sources. JPMorgan Chase (JPMC) is leading this trend by leveraging knowledge graphs across the organization for multiple mission critical applications such as risk assessment, fraud detection, investment advice, etc. A core problem in leveraging a knowledge graph is to link mentions (e.g., company names) that are encountered in textual sources to entities in the knowledge graph. Although several techniques exist for entity linking, they are tuned for entities that exist in Wikipedia, and fail to generalize for the entities that are of interest to an enterprise. In this paper, we propose a novel end-to-end neural entity linking model (JEL) that uses minimal context information and a margin loss to generate entity embeddings, and a Wide \& Deep Learning model to match character and semantic information respectively. We show that JEL achieves the state-of-the-art performance to link mentions of company names in financial news with entities in our knowledge graph. We report on our efforts to deploy this model in the company-wide system to generate alerts in response to financial news. The methodology used for JEL is directly applicable and usable by other enterprises who need entity linking solutions for data that are unique to their respective situations.
\end{abstract}

\section{Introduction}
\begin{figure*}[h]
\centering 
\includegraphics[width=0.75\textwidth]{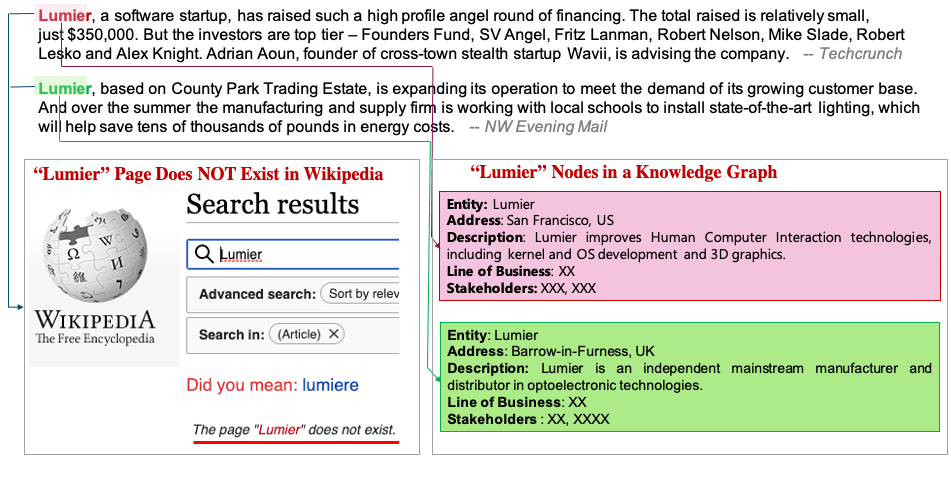}
 \caption{Example for Entity Linking}
\label{el_demo}
\end{figure*}

\emph {Knowledge Graphs} are being used for a wide range of applications from space, journalism, biomedicine to entertainment, network security, and pharmaceuticals. Within JP Morgan Chase (JPMC), we are leveraging knowledge graphs for financial applications such as risk management, supply chain analysis, strategy implementation, fraud detection, investment advice, etc. While leveraging a knowledge graph, \emph{Entity Linking} (EL) is a central task for semantic text understanding and information extraction. As defined in many studies \cite{zhang2010entity,eshel2017named,kolitsas2018end}, in an EL task we link a potentially ambiguous \emph{Mention} (such as a company name) with its corresponding \emph{Entity} in a knowledge graph. EL can facilitate several knowledge graph applications, for example,  the mentions of company names in the news are inherently ambiguous, and by relating such mentions with an internal knowledge graph, we can generate valuable alerts for financial analysts. In Figure \ref{el_demo}, we show a concrete example in which the name ``Lumier" has been mentioned in two different news items. ``Lumier"s are two different companies in the real world, and their positive financial activities should be brought to the attention of different stakeholders. With a successful EL engine, these two mentions of ``Lumier"s can be distinguished and linked to their corresponding entities in a knowledge graph.

Prior work on EL has been driven by a number of standard datasets, such as CoNLYAGO \cite{yago}, TAC KBP\footnote{https://www.ldc.upenn.edu/collaborations/current-projects/tac-kbp}, DBpedia\footnote{https://wiki.dbpedia.org/develop/datasets}, and ACE\footnote{https://catalog.ldc.upenn.edu/LDC2006T06}. These datasets are based on Wikipedia, and are therefore,  naturally coherent, well-structured and rich in context \cite{eshel2017named}.  We face the following problems when we use these  methods for entity linking for our internal knowledge graph:

\begin{enumerate}[label=\arabic*)]
\item Wikipedia does not cover all the entities of financial interest. For example, as of this writing, the startup ``Lumier" mentioned in Figure \ref{el_demo} is not present in Wikipedia, but it is of high financial interest as it has raised critical investment from famous investors.
\item Lack of context information. Many pre-trained models achieve great performance by leveraging rich context data from Wikipedia \cite{ganea2017deep}. For JPMC internal data, we do not have information comparable to  Wikipedia to support re-training or fine-tuning of existing models.
\end{enumerate}
To address the problems identified above, we built a novel entity linking system, JEL, to link mentions of company names in text to entities in our own knowledge graph. Our model makes the following advancements on the current state-of-the-art: 
\begin{enumerate}[label=\arabic*)]
\item We do not rely on Wikipedia to generate entity embeddings. With minimum context information, we compute entity embeddings by training  a \emph{Margin Loss} function.
\item We deploy the \emph{Wide \& Deep Learning} \cite{cheng2016wide} to match character and semantic information respectively. Unlike other deep learning models \cite{martins2019joint,kolitsas2018end,ganea2017deep}, JEL applies a simple linear layer to learn character patterns, making the model  more efficient both in the training phase and inference phase.
\end{enumerate}

\section{Problem Definition and Related Work}
\subsection{Problem Definition}
We assume a knowledge graph (KG) has a set of entities  $E$. We further assume that $W$ is the vocabulary of words in the input documents.  An input document $D$ is given as a sequence of words: $D=\{w_{1},w_{2},...,w_{d} \}$ where $w_{k} \in W, 1  \le k \le d$. The output of an EL model is a list  of  $T$  mention-entity pairs $\{ (m_{i},e_{i}) \}_{i \in \{1,T\}}$, where each mention is a word subsequence of $D$, $m_{i}=w_{l},...,w_{r}, 1 \le l \le r \le d$, and each entity $e_{i} \in E$. The entity linking process involves the following two steps \cite{ceccarelli2013learning}. 
\begin{enumerate}[label=\arabic*)]
\item {\bf Recognition}. Recognize a list of mentions $m_{i}$ as a set of all contiguous sequential words occurring in $D$ that might mention some entity $e_{i} \in E$. We adopted spaCy\footnote{https://spacy.io/} for mention recognition.
\item {\bf Linking}. Given a mention $m_i$, and the set of candidate entities, $C(m_i)$ such that $|C(m_{i})|>1$, from the KG, choose the correct entity, $ e_{i} \in C(m_i)$, to which the mention should be linked. We focus on solving the linking problem in this paper.
\end{enumerate}
\subsection{Popular Methods}
Entity Linking is a classical NLP problem for which the following techniques have been used: \emph{String Matching}, \emph{Context Similarity}, \emph{Machine Learning Classification}, \emph{Learning to Rank}, and \emph{Deep Learning}. In the following several paragraphs, we will briefly discuss each of them.
\begin{itemize}
\item[--] \textbf{String Matching Methods.}
String matching measures the similarity between the mention string and entity name string. We experimented with different string matching methods for name matching, including Jaccard, Levenshtein, Ratcliff-Obershelp, Jaro Winkler, and N-Gram Cosine Simiarity, and found that n-gram cosine similarity  achieves the best performance on our internal data. However, pure string-matching methods breakdown when two different entities share similar or the same name (as shown in Figure \ref{el_demo}) which motivates the need for better matching techniques.
\item[--] \textbf{Context Similarity Methods.}
Context Similarity methods compare similarities of respective context words for mentions and entities. The context words for a mention are the words surrounding it in the document. The context words for an entity are the words describing it in the KG. Similarity functions, such as Cosine Similarity or Jaccard Similarity, are widely used to compare the two sets of context words \cite{cucerzan2007large,mihalcea2007wikify}, and then to decide whether a mention and an entity should be linked. 
\item[--] \textbf{Machine Learning Classification.}
Many studies adopt machine learning techniques for the EL task. Binary classifiers, such as Naive Bayes \cite{varma2009iiit}, C4.5 \cite{milne2008learning}, Binary Logistic classifier \cite{han2011collective}, and Support Vector Machines (SVM) \cite{zhang2010entity}, can be trained on mention-entity pairs to decide whether they should be  linked.
\item[--] \textbf{Learn to Rank Methods.}
As a classification method will generate more than one  mention-entity pairs, many systems use a ranking model \cite{zheng2010learning} to select the most likely match. Learning to Rank (LTR) is a class of techniques that supplies supervised machine learning to solve ranking problems.
\item[--] \textbf{Deep Learning Methods.}
Deep learning has achieved success on numerous tasks including EL \cite{sun2015modeling,huang2015leveraging,francis2016capturing}. 
One specific model  \cite{kolitsas2018end} uses two levels of Bi-LSTM to embed characters into words, and words into mentions, and calculates the similarity between a mention vector and a pre-trained entity vector \cite{ganea2017deep} to decide whether they match. 
\end{itemize}

\section{Proposed Framework}
\subsection{Entity Embedding}
\begin{figure*}[h]
\centering 
\includegraphics[width=0.90\textwidth]{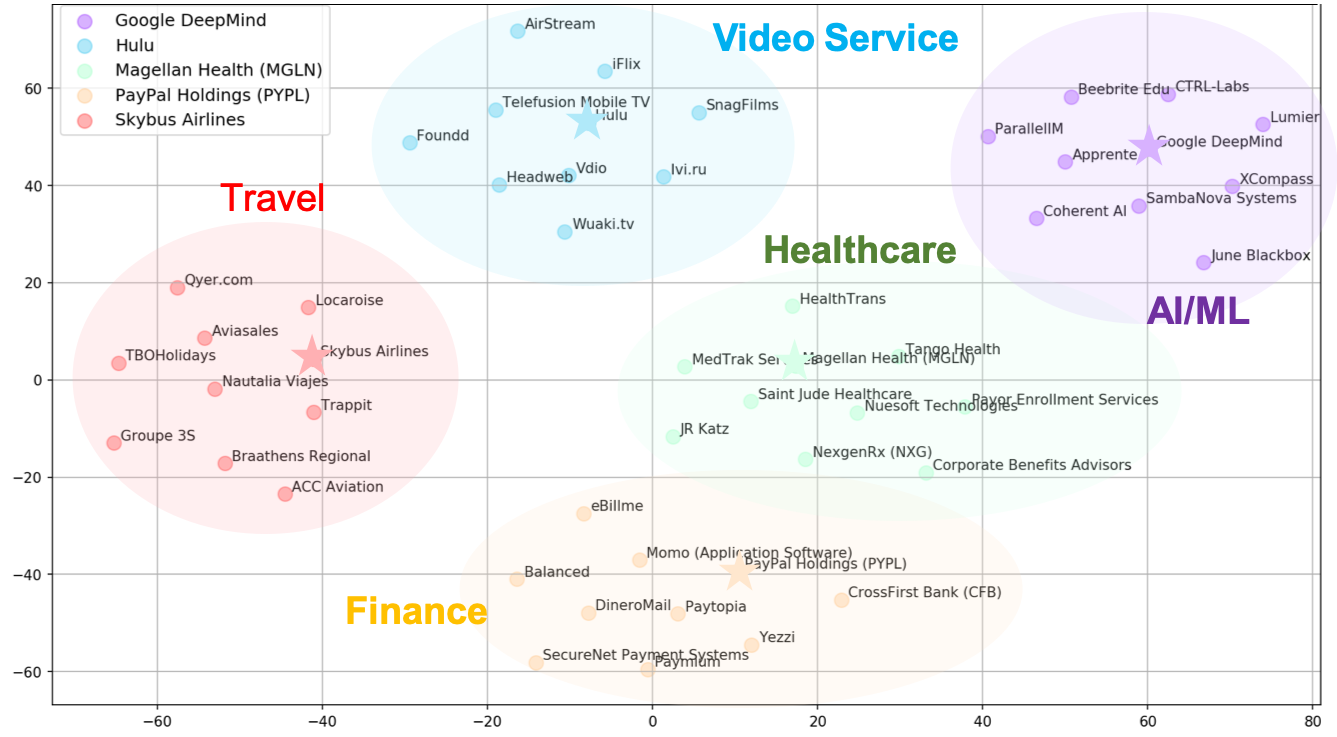}
 \caption{Visualization and Validation of Entity Embeddings with t-SNE}
\label{entity_embedding}
\end{figure*}
\label{section:entity_embedding}
Most public entity embedding models \cite{he2013learning,yamada2016joint,ganea2017deep} are designed for Wikipedia pages and require rich entity description information. In our case, each entity has a short description that is insufficient to support a solid statistical estimation of entity embeddings \cite{mikolov2013linguistic}.  To address this limitation, we use a \emph{Triplet Loss} model to generate our own entity embeddings from pre-trained word embedding models with limited context information support. 

\subsubsection{Entity Embedding Model.}
\begin{figure}[h]
\centering 
\includegraphics[width=0.45\textwidth]{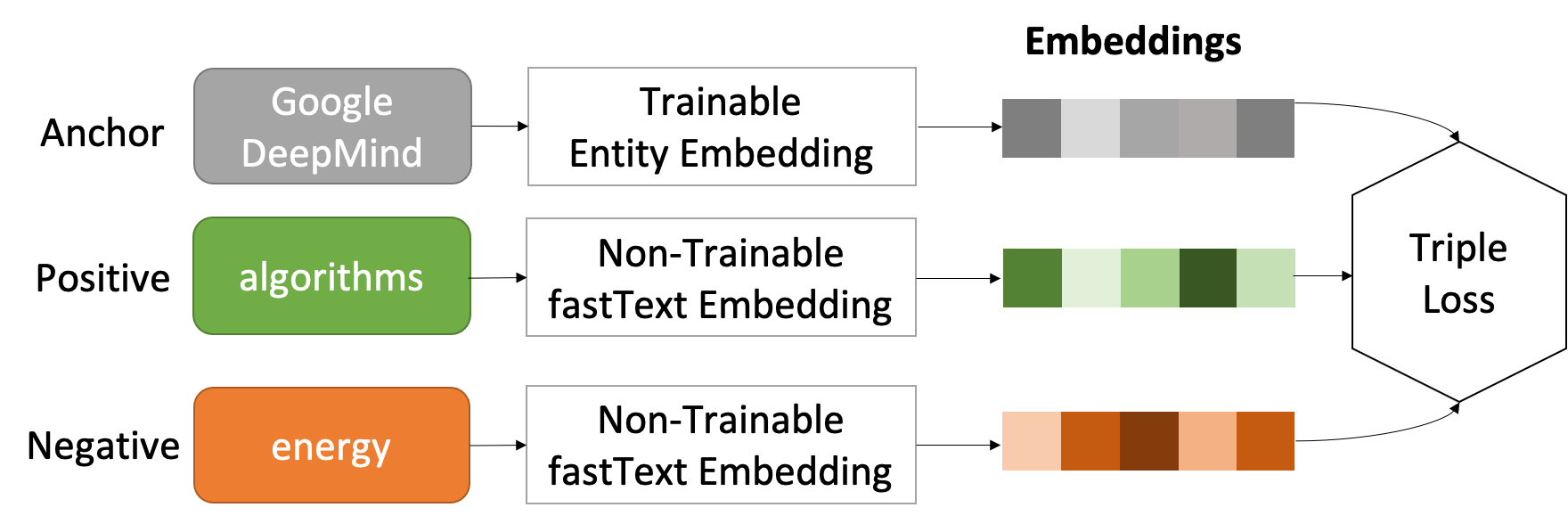}
 \caption{Model for Entity Embedding}
\label{fg:triple_loss_model}
\end{figure}

To prepare training data for this model, we select 10 words that can be used as positive examples and 10 words that can be used as negative example for each entity. To select the positive examples, we score each entity's description words with tf-idf, and select the words with 10 highest scores. To select the  negative examples, we randomly select  from words that do not appear in this entity's description.  Thus, for each entity, we can construct 10 $<$ entity, positive-word, negative-word $>$ triplets to feed into triplet loss function formulated as Equation \ref{eq:triplet_loss} below.
\begin{equation}
Loss = \sum_{i=1}^{N}[||f_{i}^{a}-f_{i}^{p}||_{2}^{2} - ||f_{i}^{a}-f_{i}^{n}||_{2}^{2}+\alpha]_{+}
\label{eq:triplet_loss}
\end{equation}
where $f_{i}^{a}$ is the vector of an anchor that we learn, $f_{i}^{p}$ is the vector from a positive sample, and $f_{i}^{n}$ is the vector from a negative sample, $\alpha$ is the margin hyper-parameter to be manually defined.  We train the entity embedding vectors ($f^{a}$). We use off-the-shelf word embedding vectors($f^{p}$ and $f^{n}$) from the fastText language model. In our experiments,  $\alpha=2.0$ led to the best performance. 

\subsubsection{Entity Embedding Validation.}
To validate the entity embeddings, we choose five seed companies from different industries --- ``Google DeepMind", ``Hulu", ``Magellan Health", ``PayPal Holdings", ``Skybus Airlines". We next select their ten nearest neighbors  (as shown in Figure \ref{entity_embedding}). We calculate a t-Distributed Stochastic Neighbor Embedding (t-SNE) to project the embeddings into a 2-dimension space. As clearly shown in Figure \ref{entity_embedding}, five seed companies from different industries are clearly separated in space. For ``Google DeepMind", we can find that all its neighbors \footnote{SambaNova Systems, Lumier, ParallelM, CTRL-labs,  Coherent AI, etc.} are, as expected,  Artificial Intelligence and Machine Learning companies. This visualization gives us a sanity check for our  entity embeddings. 

\subsection{Entity Linking}
\label{section:entity_linking}
Two factors affect an EL model's performance: \emph{Characters} and \emph{Semantics}. 
\begin{itemize}
\item[--] \textbf{Characters}: ``Lumier" will be easily distinguished from `` ParallelM" because they have completely different character patterns. These patterns can be easily captured by a wide and shallow linear model.  
\item[--] \textbf{Semantics}. In Figure \ref{el_demo}, ``Lumier(Software)" can be distinguished from ``Lumier (LED)" because they have different semantic meanings behind the same name. These semantic differences can be captured by a deep learning model. 
\end{itemize}

\begin{figure*}[h]
\centering 
\includegraphics[width=0.75\textwidth]{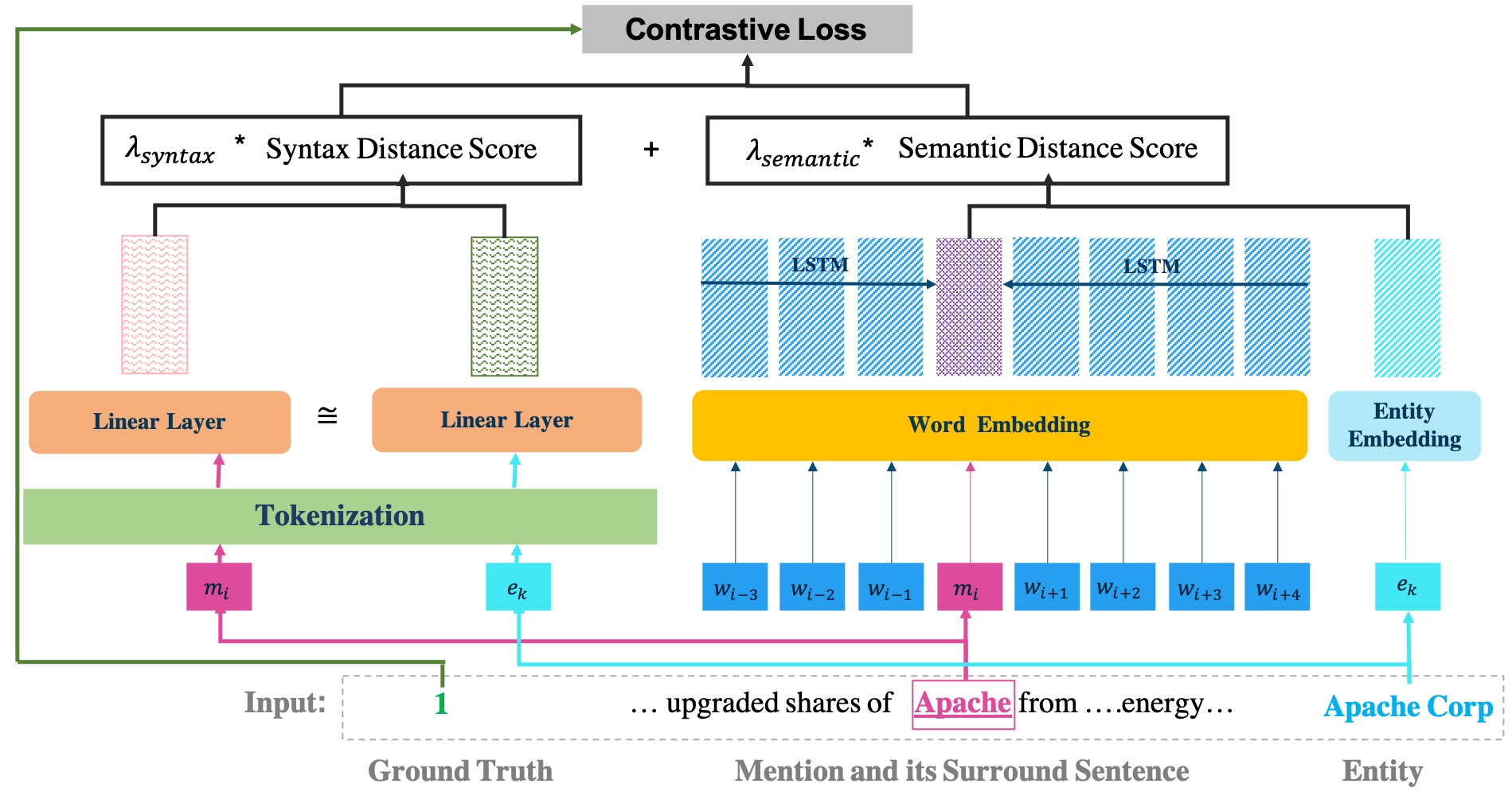}
 \caption{JEL Model Framework}
\label{fg:model_overview}
\end{figure*}

To combine the two important factors listed above, we develop a Wide\&Deep Learning model \cite{cheng2016wide} for our EL task (shown in Figure \ref{fg:model_overview}).

\subsubsection{Wide Character Learning.}
Unlike many other approaches \cite{kolitsas2018end,lample2016neural} that apply character embeddings to incorporate lexical information,  we apply a wide but shallow linear layer for the following two reasons. First, embedding aims to capture an item's semantic meanings, but characters naturally have no such semantics. ``A" in ``Amazon" does not have any relationship with ``A" in ``Apple".  Second, as embedding layer involves more parameters to optimize, it is much slower in training and inference than a simple linear layer. 

\textbf{Feature Engineering.}
Many mentions of an entity exhibit a complex morphological structure that is hard to account for by simple word-to-word or character-to-character matching. Subwords can improve matching accuracy dramatically. Given a string, we undertake the following processing  to maximize morphological information we can get from subwords.  
\begin{enumerate}[label=\arabic*)]
\item Clean a string, convert it to lower case, remove punctuation, standardize suffix, etc. For example, ``PayPal Holdings, Inc. " will change to ``paypalhlds". 
\item Pad the start and end of the string;``paypalhlds" will be converted to ``*paypalhlds*".
\item Apply multiple levels of n-gram ($n \in [2,5]$) segmentation;``*paylpalhlds* " will be \{ *p, ay,..., lhlds, hlds* \}.
\item Append original words, *paypal* and *hlds*, to the token list.
\end{enumerate}

\textbf{Wide Character Learning.}
We applied a \emph{Linear Siamese Network} \cite{bromley1994signature} for wide character learning. We implemented two identical linear layers with shared weights (as shown in the left part of Figure \ref{fg:model_overview}). With this architecture, similar inputs, $T_{m}$ and $T_{e}$, will generate similar outputs, $Y_{m}$ and $Y_{e}$. We applied the Euclidean distance to estimate output's similarity. 
\begin{equation}
\label{syntax_distance}
D_{syx} = d(\mathbf{Y_{m}}, \mathbf{Y_{e}}) = \sqrt{\sum_{i=1}^{n}(Y_{m_{i}}-Y_{e_{i}})^{2}}
\end{equation}

\subsubsection{Deep Semantic Embedding.}
We embed the mentions into vectors so that we can mathematically measure similarities between them and entity embeddings. We use LSTM to embed mentions from their context. Similar idea is adopted elsewhere \cite{kolitsas2018end}, but instead of using a Bi-LSTM over the whole context, we apply two shorter LSTMs to embed mention  from two directions (as shown in Figure \ref{fg:model_overview}), making the embedding more targeted with less parameters involved. 

Given a mention $m_{t}$, we treat its left $n$ words $\{w_{t-n},..., w_{t-1}, w_{t} \}$ (mention words included) and its left context, and right $n$ words$\{w_{t}, w_{t+1},..., w_{t+n}\}$ as its right context (mention words included)\footnote{We adopt the same pre-trained word embedding as for entity embedding.}
\begin{equation}
\begin{split}
h_{t}^{l} = \overrightarrow{LSTM}(w_{t-1}^{l},w_{t})\\
h_{t}^{r} = \overleftarrow{LSTM}(w_{t+1}^{r},w_{t})
\end{split}
\end{equation} 
In addition to LSTM, we apply an attention layer to distinguish the influence of words. We multiply last layer's output from LSTM $\{x_{i},..,x_{j}\}$ with attention weights, and get a context representation $v$.
\begin{equation}
\begin{split}
\alpha_{k}=<\mathbf{w}_{\alpha},x_{k}> \\
a_{k} = \frac{exp(\alpha_{k})}{\sum_{s=i}^{j}exp(\alpha_{s})}\\
g = \sum_{k=i}^{j}a_{k}x_{k}
\end{split}
\end{equation}
Thus, we form a mention's vector by concatenating its left and right context representations, $g_{l}$ and  $g_{r}$:
\begin{equation}
\begin{split}
g_{m}=[g_{l};g_{r}]\\
V_{m}=FC(g_{m})
\end{split}
\end{equation}
where $FC$ is a fully connected feed-forward neural network. 
When we get the mention embedding $V_{m}$, given a pretrained entity embedding vector $V_{e}$, we can calculate similarity between these two vectors based on  Euclidean distance.
\begin{equation}
\label{semantic_distance}
D_{smc} = d(\mathbf{V_{m}}, \mathbf{V_{e}}) = \sqrt{\sum_{i=1}^{n}(V_{m_{i}}-V_{e_{i}})^{2}}
\end{equation}
\subsubsection{Contrastive Loss Function}
We combine both $D_{syx}$ and $D_{smc}$ as our target to train the model. The final distance is defined as:
\begin{equation}
D_{W} = \lambda_{syx} D_{syx} +\lambda_{smc} D_{smc}
\end{equation}
Then, we apply a contrastive loss function to formulate our object loss function. 
\begin{equation}
L=(Y) \frac{1}{2}(D_{W})^2 +(1-Y) \frac{1}{2} \{max (0,m-D_{W})\}^2
\end{equation}
where $Y$ is the ground truth value, where a value of  $1$ indicates that mention $m$ and entity $e$ is matched, $0$ otherwise.

\section{Experiment and Analysis}
\begin{table*}[h]
\footnotesize
\centering
\begin{tabular}{|c|c|c|c|c|c|c|c|c|}
\hline
                & True Positivve & True Negative & False Positive & False Negative & Precision & Recall & F1-Sccore & Accuracy \\ \hline
JEL             & 0.5            & 0.4991        & 0.0009         & 0              & 0.9982    & 1      & 0.9991    & 0.9991   \\ \hline
SVM-Rank        & 0.4826         & 0.4826        & 0.0174         & 0.0174         & 0.9652    & 0.9652 & 0.9652    & 0.9652   \\ \hline
ENEL            & 0.3875         & 0.4922        & 0.0078         & 0.1125         & 0.9803    & 0.775  & 0.8656    & 0.8797   \\ \hline
Tri-Gram Cosine & 0.4423         & 0.4303        & 0.0666         & 0.0577         & 0.8691    & 0.8846 & 0.8768    & 0.8726   \\ \hline
LR              & 0.4942         & 0.3712        & 0.1288         & 0.0058         & 0.7933    & 0.9884 & 0.8801    & 0.8654   \\ \hline
SVM             & 0.4958         & 0.3589        & 0.1411         & 0.0042         & 0.7785    & 0.9916 & 0.8722    & 0.8547   \\ \hline
Bi-Gram Cosine  & 0.4496         & 0.3794        & 0.1206         & 0.0504         & 0.7885    & 0.8992 & 0.8402    & 0.829    \\ \hline
Jaccard Context & 0.2444         & 0.3643        & 0.1357         & 0.2556         & 0.6430    & 0.4888 & 0.5554    & 0.6087   \\ \hline
Cosine Context  & 0.4912         & 0.0116        & 0.4885         & 0.0088         & 0.5014    & 0.9824 & 0.6639    & 0.5028   \\ \hline
\end{tabular}
\caption{Performance Comparison via Precision and Recall}
\label{tb:accuracy}
\end{table*}

\subsection{Data Preparation}
We first applied spaCy over financial news to detect all the named entity mentions. SpaCy features neural models for named entity recognition (NER).  By considering text capitalization and context information, spaCy claims an accuracy above 85\% for NER. Satisfied with spaCy's performance, we used  it on financial news to recognize all the critical mentions that are tagged with ``ORG". The  data preparation process was as follows:

\begin{enumerate}[label=\arabic*)]
\item We extracted mentions from the financial news with spaCy. 
\item We applied bi-gram cosine similarity between the extracted mentions and company names in our internal knowledge grpah. 
\item  If the similarity score between a mention string and an entity name is smaller than 0.5, we treated that as a strong signal that the two are not linked, and marked them as 0.
\item If the similarity score between a mention string and an entity name is equal to 1.0, we manually checked the list to avoid instances that two different entities share the same name (infrequent), and marked the pair as 1.
\item If a mention and an entity name have cosine similarity larger than 0.75, but smaller than 1.0, we manually labeled: 
\begin{enumerate}
\item Some cases are easy to tell, such as: “Luminet” vs “Luminex”, we labeled those instances as 0 directly. 
\item Some cases can be decided according to their description/context. We printed mention’s context information and entity's description respectively, and made the decision based on those texts, such as ``Apple" vs ``Apple Corps.". 
\item Some other cases need help  from publicly information found through internet search to decide, such as  "Apollo Management" vs "Apollo Global Management".
\end{enumerate}
\item If a mention and an entity name have cosine similarity between 0.5 and 0.75, we discarded it. These cases are too many for manual labeling, and too complicated for machine labeling.
\item Negative examples from step 3, make the dataset very imbalanced containing many more negative pairs. We counted the number of examples obtained in steps 4. and 5., and randomly sampled a comparable number from the examples gathered in step 3.
\end{enumerate}

In total, we have labeled 586, 975 ground truth mention-entity pairs, with 293,949 positive mention-entity pairs, and 293, 026 negative pairs. We split 80\% of the data as training data, 10\% as validation data, and 10\% as testing data.

\subsection{Baselines}
\begin{enumerate}[label=\arabic*)]
\item \textbf{String Matching} 
We chose Bi-Gram and Tri-Gram Cosine Similarity  as two of baselines. Before similarity calculation, all tokens were weighted with tf-idf scores. We set 0.8 as the threshold. 
\item \textbf{Context Similarity}
We used Jaccard and Cosine similarity to measure the similarities between mention context and entity descriptions.  A potential matched mention-entity pair should share at least one context word. 
\item \textbf{Classification}
We chose Logistic Regression (LR) and SVM for experiments. We adopted the feature engineering method defined in \cite{zheng2010learning}, but only kept the following features that we can generate from our data:
\begin{itemize}
\item[--] StrSimSurface: edit-distance among mention strings and entity names. 
\item[--] ExactEqualSurface: number of overlapped lemmatized words in mention  strings and entity names. 
\item[--] TFSimContext: TF-IDF similarity between mention's context and entity's description
\item[--] WordNumMatch: the number of overlapped lemmatized words between mention's context and entity's description.
\end{itemize}

\item \textbf{Learn to Rank}
We used SVM-RANK as the representation of Learn to Rank. We adopted the same features as defined above. 
\item \textbf{Deep Learning}
We implemented state-of-the-art deep learning algorithm \cite{kolitsas2018end} (ENEL) as one of our baselines. In their original study, the authors utilized a Wikipedia derived entity embedding. But we utilized our own entity embedding for learning.
\end{enumerate}
\subsection{Comparison on Accuracy}
We first compare the methods with \emph{Precision} and \emph{Recall}. For an easier comparison, we  scaled each of \emph{True Positive}, \emph{True Negative}, \emph{False Positive}, and \emph{False Negative} into [0,0.5] showing as following. 
\begin{itemize}
\item[--] True Positive=$\dfrac{\text{Count(Predict=1\ \&\ Truth=1)}}{2 \times \text{Count(Truth=1)}}$\\
\item[--] True Negative=$\dfrac{\text{Count(Predict=0\ \&\ Truth=0)}}{2 \times \text{Count(Truth=0)}}$\\
\item[--]  False Positive=$\dfrac{\text{Count(Predict=1\ \&\ Truth=0)}}{2 \times \text{Count(Truth=0)}}$\\
\item[--] False Negative=$\dfrac{\text{Count(Predict=0\ \&\ Truth=1)}}{2 \times \text{Count(Truth=1)}}$
\end{itemize}

The result is shown as Table \ref{tb:accuracy}, in which:
\begin{itemize}
\item[--] Precision=$\dfrac{\text{True Positive}}{\text{True Positive}+\text{False Positive}}$\\
\item[--] Recall=$\dfrac{\text{True Positive}}{\text{True Positive}+\text{False Negative}}$\\
\item[--] F1-Score=$ 2 \times \dfrac{\text{Precision} \times \text{Recall}}{\text{Precision}+\text{Recall}}$\\
\item[--] Accuracy = True Positive+True Negative
\end{itemize}

From Table \ref{tb:accuracy}, we find context based methods perform poorly as expected. Descriptions in our knowledge graph have very different wording styles from financial news. Simply comparing context words will definitely result in low accuracy.  SVM-Rank surprisingly outperforms ENEL. The reason here is that ENEL does not model character features properly. In SVM-Rank, we have carefully designed character features, (e.g., edit distance and tf-idf similarity), but ENEL just embeds 36 single character embeddings. This result also indicates that without good character learning, even deep learning could not solve the linking problem well.

JEL performs the best. We will mainly discuss the reason that JEL outperforms ENEL. First JEL involves more character features. ENEL just embeds 36 characters (26 letters + 10 digits), but JEL computes 151622 character features (as shown in Table \ref{el-table}). This configuration supports JEL with a better performance in capturing character patterns. For example, JEL could successfully link ``Salarius Pharm LLC" to ``Salarius Pharmaceuticals" but ENEL missed this link. Second, ENEL jointly embeds all characters and words from context and mention itself into a mention's vector, and minimizes the distance between this mention vector and a pre-trained entity embedding vector. However, the entity embeddings  themselves are generated without character information \cite{ganea2017deep}. Character embeddings in ENEL, especially character embeddings from context words, somehow add noise to semantic embeddings, and impact final performance. In addition, Table \ref{el-table} gives a brief overview of efficiency comparison between JEL and ENEL. Although JEL and ENEL share similar number of parameters, JEL trains faster than ENEL. JEL utilizes linear layers to learn character patterns, which is easier to learn than an embedding layer in ENEL. 
\begin{table}
\footnotesize
\centering
\begin{tabular}{|c|c|c|c|}
\hline
        & \#parameters & \begin{tabular}[c]{@{}c@{}}\# character \\ features\end{tabular} & \begin{tabular}[c]{@{}c@{}}processing \\ time\end{tabular} \\ \hline
JEL &  10915800    & 151622                                                           & $\sim$1min/batch                                           \\ \hline
ENEL &  11021400    & 36                                                               & $\sim$20 min/batch                                         \\ \hline
\end{tabular}
\caption{Comparison among JEL and ENEL}
\label{el-table}
\end{table}

\subsection{Comparison on Precision}
\begin{figure}
\centering 
\includegraphics[width=0.45\textwidth]{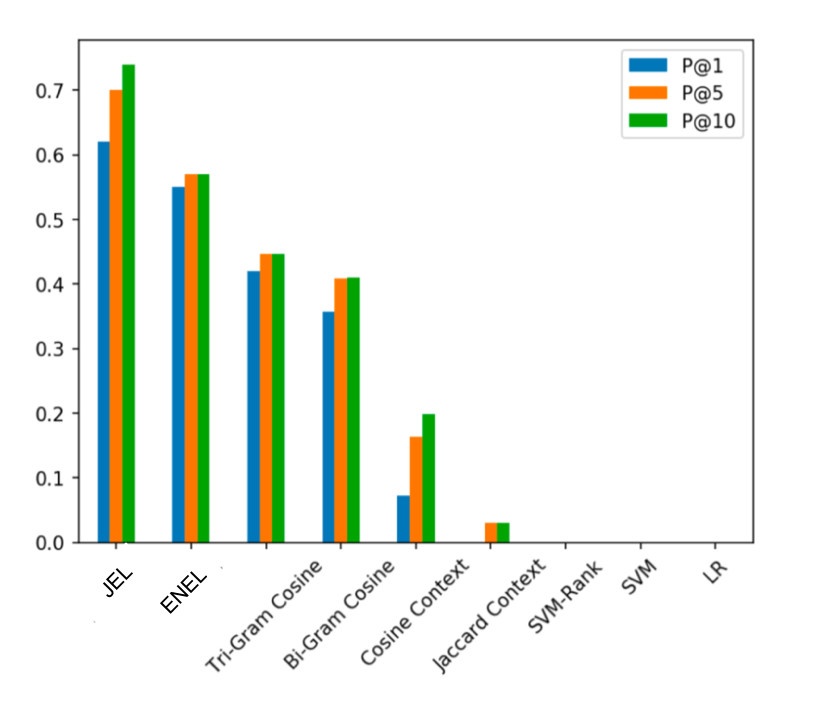}
 \caption{Performance Comparison via P@K(Precision at top K)}
 \label{precision}
\end{figure}

Accuracy can only check a method's ability in distinguishing positive samples from negative samples. In a real EL task, we care more about a method's ability in finding the correct entity for a given mention.  In this section, we utilize the ``Precision at top K (P@K)" to compare the methods (shown in Figure \ref{precision} with $K \in \{1,5,10\}$). LR, SVM, and SVM-Rank get 0s at all P@Ks. Both LR and SVM are classifiers, which are good at distinguishing positive pairs from negative pairs. However, they may label multiple mention-entity pairs as positive, but rank false positive pairs ahead. The SVM-Rank's failure is surprising. We believe the reason lies in the training data. We only have $1$ and $0$ as labels, but no specific ranking order, which is hard to obtain for a $0-1$ problem, to train the model solidly. Without a ranking ground truth, SVM-Rank fails in our task. For the other four methods, we rank entities based on distance or similarity. JEL still achieves the best performance.

\section{Deployment and Business Impact}
JPMC has built a large-scale knowledge graph for internal use, that integrates data from third party providers with its internal data. The system contains several million entities (e.g. suppliers, investors, etc.) and several million links (e.g. supply chain, investment, etc.) among those entities \footnote{For proprietary reasons, we cannnot reveal the exact number of nodes and links}. Entity linking is one of the core problems that needs to be solved when ingesting unstructured data. Currently, we are working closely with business units to process incoming news articles to extract news about companies, and connect them to the corresponding entities in the  knowledge graph. Such news analytics will 
support users across JPMC in discovering relevant and curated news that matter to their business. 

Here we present a concrete example of the use of such news analytics. To protect the customer cofidentiality, the actual name of the company has been changed, but the illustrated computation is the same. ``Acma Retail Inc" filed for bankruptcy due to the pandemic, and a lot of JPMC clients could feel stress as they are suppliers to Acma. Such stress can pass deep down into its supply chain and trigger financial difficulties for other clients. JPMC may face different levels of risks from suppliers with different orders in Acma’s supply chain.  With ``Acma" mentioned in financial news linked with ``Acma Global Retail Inc" in our knowledge graph (distinguished from ``Acma Furniture, LLC", ``Acma Enterprise System", etc.), we can accurately track down Acma supply chain, identify stressed suppliers with different revenue exposure, and measure our primary risk due to Acma’s bankruptcy. Once stressed clients with significant exposure are detected, alerts will be sent out to corresponding credit officers. If ``Acma" was linked with incorrect entities, it will result in too many false signals, resulting in wasted effort.

A similar news analytics system, SNOAR, has been previously reported \cite{goldberg2003nasd}. SNOAR used a rule based and fuzzy match techniques for entity linking, which are not able to handle the example illustrated in Figure \ref{el_demo}. In fact, before JEL was developed, \emph{Tri-Gram Cosine Similarity} was tested for name string matching. As shown in Table \ref{tb:accuracy} and Figure \ref{precision}, tri-gram cosine similarity has practical limitations and could not distinguish entities sharing similar names (as shown in Figure \ref{el_demo}). Compared to fuzzy match  or tri-gram cosine similarity, JEL provides a better and more controllable solution.

In the deployed version of JEL, we will apply another blocking layer (overlap blocker for current configuration) ahead of JEL to reduce the candidate volume for each mention in an article. Entities sharing less than 2 bi-gram tokens will be filtered out in the blocking stage, and the rest of candidate entities will be sent to JEL for a more sophisticated linking proccess. This blocking process will dramtically reduce computational cost. 

During the first-round deployment, we are incorporating JEL into a streaming news platform, and its online performance will be tested on indicators including: 
\begin{itemize}
\item[--] \emph{Scope}: Ability in linking all available JPMC clients 
\item[--] \emph{Timeliness}: Ability in providing near real-time alerts 
\item[--] \emph{Accuracy}: Ability in sending accurate alerts 
\item[--] \emph{Flexibility}: Ability in integration of various components for end-to-end delivery of solutions. 
\end{itemize}

Our first step is to collect user feedbacks for parameter fine-tuning and algorithm re-configuration. Once the initial deployment is successful, we will release JEL as a standalone and reusable component with an API to provide a firm-wide service that can be used in various applications.

\section{Conclusion}
Knowledge Graphs are becoming a mission critical technology across many industries. Within JPMorgan Chase, we
are using a knowledge graph as a company-wide resource for tasks such as 
risk analysis, supply chain analysis, etc.  A core problem in utilizing this knowledge is EL. 
Existing EL models did not generalize to our internal company data, and therefore, we developed a novel
model, JEL, that leverages margin loss function and deep and wide learning. Through an extensive 
experimentation, we have shown the superiority of this method. We are currently in the process of deploying 
this model on a financial news platform within our company. Even though our testing was done on 
our internal data, we believe, that our approach can be adapted by other companies for entity linking
tasks on their internal data.
\bibliography{reference}
\end{document}